\documentclass[12pt]{article}

\usepackage{algorithm}
\usepackage{algpseudocode}
\usepackage{amsmath}
\usepackage{amssymb}
\usepackage{harvard}
\usepackage{hyperref}
\usepackage{graphicx}
\usepackage{lastpage}

\makeatletter
\newenvironment{breakablealgorithm} {
    	\begin{center}
	\refstepcounter{algorithm}
        	\renewcommand{\caption}[1] {
		\addcontentsline{loa}{algorithm}{\protect\numberline{\thealgorithm}##1}
            	\parbox{\textwidth}
            	{
                		\hrule height.8pt depth0pt \kern2pt
                		{\raggedright\textbf{\fname@algorithm~\thealgorithm} ##1\par}
                		\kern2pt\hrule\kern2pt
            	}
        	}
}
{
	\kern2pt\hrule\relax
	\end{center}
}
\makeatother

\title{Comprehensive Stepwise Selection for Logistic Regression}

\author
{Bernd Engelmann$^{1\ast}$\\
\\
\normalsize{$^{1}$Department of Finance \& Banking, Ho Chi Minh City Open University,}\\
\normalsize{35 - 37 Ho Hao Hon, District 1, Ho Chi Minh City, Vietnam}\\
\\
\normalsize{$^\ast$E-mail: bernd.engelmann@ou.edu.vn}
}

\date{\today}

\begin{document}

\baselineskip16pt

\maketitle

\begin{abstract}
Automated variable selection is widely applied in statistical model development. Algorithms like forward, backward or stepwise selection are available in statistical software packages like R and SAS. Many researchers have criticized the use of these algorithms because the models resulting from automated selection algorithms are not based on theory and tend to be unstable. Furthermore, simulation studies have shown that they often select incorrect variables due to random effects which makes these model building strategies unreliable. In this article, a comprehensive stepwise selection algorithm tailored to logistic regression is proposed. It uses multiple criteria in variable selection instead of relying on one single measure only, like a $p$-value or Akaike's information criterion, which ensures robustness and soundness of the final outcome. The result of the selection process might not be unambiguous. It might select multiple models that could be considered as statistically equivalent. A simulation study demonstrates the superiority of the proposed variable selection method over available alternatives.
\end{abstract}

\section{Introduction}

Automated variable selection algorithms for regression models have been widely applied in various areas of research \cite{harrell2015regression,heinze2018variable}. The most commonly used methods are forward, backward and stepwise variable selection. The idea of forward selection is starting with a constant model and add variables one-by-one. The criterion to add variables could be statistical significance of model coefficients, i.e., add the variable with the lowest $p$-value, or applying an information measure like Akaike's Information Criterion (AIC) or the Baysian Information Criterion (BIC). Backward selection works the other way round. Here, the starting point is a model containing all variables and the least important variables are removed one-by-one until a stopping criterion is reached. Stepwise selection is an extension of forward selection which is more sophisticated as it allows for the removal of variables in later selection steps.\\

The application of automated selection algorithms has been widely criticized. \citeasnoun{smith2018step} demonstrated by means of a simulation study that stepwise selection algorithms have fundamental problems that cannot be attributed to lack of data but even occur with big data. Although \citeasnoun{smith2018step} is relatively recent research, concerns about misleading outcome of automated selection algorithms are not new and have been been raised in multiple studies over the past decades \cite{austin2004automated,flom2007stopping,whittingham2006we}. Proposals to overcome the shortcomings of automated model selection have been made. One possibility is combining automated selection with cross validation to improve the control of potential model instabilities \cite{harrell2015regression,heinze2018variable}. A popular alternative is penalizing the size of model coefficients during model estimation, effectively reducing the number of variables included in a model. Depending on the shape of the penalization term, these methods are known as Ridge regression \cite{schaefer1984ridge,le1992ridge}, the Lasso \cite{meier2008group}, or the elastic net which is essentially a combination of both \cite{zou2005regularization}. These methods are useful in preventing overfitting. However, they cannot solve the fundamental problems outlined in \citeasnoun{smith2018step} like the lack of theoretical consideration in the model building process. Furthermore, as will be demonstrated in this article, they only mitigate shortcomings of simple forward, backward and stepwise selection but do not eliminate them.\\

Despite of the known problems with existing automated model selection algorithms, there are practical applications where a large number of regression models has to be estimated which requires automation. An example could be the estimation of credit risk models for multiple countries and asset classes for investors in loan portfolios. Having an automated selection algorithm that is able to identify sensible well-functioning models would be a valuable support for a data analyst facing this problem. In this article, an automated selection algorithm is proposed for logistic regression, one of the most popular statistical modeling techniques that is applied in many scientific areas. Contrary to the aforementioned selection algorithms which are quite generic and could be applied to various families of regression models, the framework developed in this article will be tailored to logistic regression and cannot be easily transferred to other model classes, like linear or multinomial regression.\\

Logistic regression is a binary classification model. Its quality can be measured in two dimensions, discrimination and calibration. By discrimination, the ability of a logistic regression model to separate good from bad observations is measured. Popular measures for this purpose are the Accuracy Ratio \cite{engelmann2003measuring} and the area below the Receiver Operating Characteristic (ROC) curve \cite{swets2014signal}. The second dimension, calibration, refers to the accuracy of probability estimates for the bad event. Calibration could be measured by the mean squared error, in the context of logistic regression also known as Brier score \cite{brier1950verification}. The stepwise selection algorithm proposed in this article will heavily rely on these two notions. The aim is selecting variables that lead to an overall improvement of discriminative power and calibration. In addition, there will be controls for statistical significance of model coefficients, multi-collinearity, model overfitting, and theoretical soundness.\\

 A key difference of the selection algorithm in this article and simple forward and backward selection is in the outcome. Forward and backward selection will by construction always return one model which is interpreted as the best model according to the criterion that is applied in the selection process. The selection algorithm in this article might deliver multiple solutions. In this case, these solutions could be considered as equivalent in a statistical sense, i.e., when tests on difference in either discrimination or calibration are applied, these tests are unable to distinguish between these models.\\
 
In the next section, the Comprehensive Stepwise Selection for Logistic Regression (CSSLR) algorithm is introduced and explained in detail. The main motivation of the algorithm is combining multiple criteria for evaluating the quality of a logistic regression model in a structured way to ensure robustness of the final outcome. Section \ref{sec:performance} will illustrate the performance of the algorithm on simulated data. The final section concludes. In the appendix, it is briefly explained how to install and run an implementation of CSSLR in R.

\section{The CSSLR Algorithm}

The CSSLR algorithm is a stepwise selection algorithm that starts with a constant model and adds variables one-by-one in every step. It allows for the removal of variables in later steps should they turn out to become irrelevant once a model is growing. The algorithm stops when it is no longer possible to improve the set of selected models by adding more variables. On a high level, one selection step of the algorithm is described below.

\begin{breakablealgorithm}\label{alg:select}
	\caption{High-Level Algorithm of a Selection Step}
	\begin{algorithmic}
		\State Let $\pmb{\mathbb{M}} = \left(\mathbb{M}_1,\ldots,\mathbb{M}_n\right)$ be the models selected in the previous steps
		\State Let ${\mathbf V} = \left(V_1,\ldots,V_m\right)$ be the set of variables contained in the data set
		\vspace*{0.2cm}
		\State {\bf Part I}: Identification of Improved Models
		\vspace*{0.1cm}
		\For{$i=1,\ldots,n$}				\Comment{Loop over all previously selected models}
			\For{$V_j,\ V_j\notin \mathbb{M}_i$}		\Comment{Loop over all variables not contained in $\mathbb{M}_i$}
				\State Estimate model $\mathbb{M}_c$ containing the variables in $\mathbb{M}_i$ and the new variable $V_j$
				\If{$\mathbb{M}_c$ is an {\bf improved} model}
					\State {\bf Trim} model $\mathbb{M}_c$ if possible and required
					\State Add model $\mathbb{M}_i$ to the set to models to be deleted $\pmb{\mathbb{D}}$
					\State Add model $\mathbb{M}_c$ to the set of improved models $\pmb{\mathbb{I}}$
				\Else
					\State Discard model $\mathbb{M}_c$
				\EndIf
			\EndFor
		\EndFor
		\vspace*{0.2cm}
		\State {\bf Part II}: Identification of Equivalent Models
		\vspace*{0.1cm}
		\If{$\pmb{\mathbb{I}}$ is empty}
			\State Stop selection algorithm and return the solution $\pmb{\mathbb{ M}}$
		\Else
			\State Remove all models in $\pmb{\mathbb{D}}$ from $\pmb{\mathbb{M}}$ and add the models in $\pmb{\mathbb{I}}$
			\State Find the {\bf leading} models in $\pmb{\mathbb{M}}$, $\mathbb{M}_1$ and $\mathbb{M}_2$ \Comment{Leading model could be unique}
			\For{$\mathbb{M}\in\pmb{\mathbb{M}}$ and $M\neq \mathbb{M}_1,\mathbb{M}_2$}
				\If{$\mathbb{M}$ is {\bf equivalent} to $\mathbb{M}_1$ and $\mathbb{M}_2$}
					\State Keep $\mathbb{M}$ in $\pmb{\mathbb{M}}$
				\Else
					\State Remove $\mathbb{M}$ from $\pmb{\mathbb{M}}$
				\EndIf
			\EndFor
		\EndIf 
	\end{algorithmic}
\end{breakablealgorithm}

Algorithm \ref{alg:select} essentially consists of two parts. In the first part variables are added one-by-one to already existing models. A set of models $\pmb{\mathbb{I}}$ is constructed which contains all models that have shown an improvement over the existing models. In the second part of a selection step the improved models are compared among each other. Models that are inferior are discarded and only a smaller set of models is carried forward to the next selection step. All models in the smaller set are considered as equivalent.\\

Algorithm \ref{alg:select} is entirely descriptive. To understand how it works on a data set, the notion of improved model, trimmed model, leading model and equivalent model has to be defined in statistical terms. In all these steps, two models are compared and various statistical quantities are computed. From the outcome it can be decided if a model is improved compared to a second model, should be trimmed, is leading among a set of models, or is equivalent to another model.\\

To introduce some notation, let $I$ be an indicator variable which is 0 when an observation in a data set is good and 1 when it is bad. Suppose, a logistic regression model $\mathbb{M}\in\pmb{\mathbb{M}}$ contains the variables $V_1,\ldots,V_m$. The model equation is
\begin{equation}\label{eq:logit}
	-\log\left(\frac{1-P\left(I=1|V_1,\ldots,V_m\right)}{P\left(I=1|V_1,\ldots,V_m\right)}\right) = \beta_0 + \sum_{i=1}^m\beta_i\cdot V_i.
\end{equation}
In the next selection step, a candidate variable $V_c$ with $V_c\notin\left(V_1,\ldots,V_m\right)$ is added to model $\mathbb{M}$ resulting in model $\mathbb{M}_c$
\begin{equation}\label{eq:logit:candidate}
	-\log\left(\frac{1-P\left(I=1|V_1,\ldots,V_m,V_c\right)}{P\left(I=1|V_1,\ldots,V_m,V_c\right)}\right) = \tilde{\beta}_0 + \sum_{i=1}^m\tilde{\beta}_i\cdot V_i + \tilde{\beta}_c\cdot V_c.
\end{equation}

To evaluate whether model $\mathbb{M}_c$ is an improvement over model $\mathbb{M}$, model $\mathbb{M}_c$ has to fulfill some minimum requirements like the statistical significance of $\tilde{\beta}_c$. In addition, it has to be better than model $\mathbb{M}$. Therefore, performance measures have to be analyzed which allow to decide whether $\mathbb{M}_c$ is an improvement over $\mathbb{M}$. This is the step that is tailored to logistic regression.\\

As outlined above, the quality of a logistic regression model can be measured in terms of discrimination and calibration. Discrimination can be measured by the area under the ROC curve (AUC). An overview of different approaches for its calculation can be found in \citeasnoun{faraggi2002estimation}. A requirement for model improvement should be $AUC\left(\mathbb{M}_c\right) > AUC(\mathbb{M})$. To make sure that this effect is not just due to data noise, a statistical test on the difference of two models' AUC should be applied \cite{delong1988comparing}. As a decision criterion, one could define a critical $p$-value for the AUC-test, $p_{AUC,I}$, and require that the $p$-value of the test comparing $AUC\left(\mathbb{M}_c\right)$ with $AUC(\mathbb{M})$ is less than $p_{AUC,I}$.\\

As a measure for calibration, the mean squared error is widely used. It is computed from estimated probabilities $\pi = P(I=1)$ for being bad and the realization of the binary variable $I$:
\begin{equation}\label{eq:mse}
	MSE = \frac{1}{N}\sum_{i=1}^N\left(\pi_i - I_i\right)^2,
\end{equation}
where $N$ is the sample size of a data set. For an improvement in calibration, the requirement is $MSE(M_c) < MSE(M)$. To make this decision statistically sound, two tests should be applied. The first test, \citeasnoun{spiegelhalter1986probabilistic}, checks whether each model individually is well calibrated. Here, the null hypothesis is that MSE is equal to its expected value $E\left[MSE\right]$ and it should not be possible to reject it. Therefore, one defines a critical $p$-value for this test $p_{calib}$ and requires that this test's $p$-value is greater than $p_{calib}$. Only, if both models $\mathbb{M}$ and $\mathbb{M}_c$ pass the test of \citeasnoun{spiegelhalter1986probabilistic}, a test for comparing $MSE(\mathbb{M}_c)$ and $MSE(\mathbb{M})$ can be performed \cite{redelmeier1991assessing}. Analogously to the $AUC$ test, a critical $p$-value $p_{MSE,I}$ is defined and to ensure that $MSE(\mathbb{M}_c)$ is below $MSE(\mathbb{M})$ with statistical significance, the $p$-value of the \citeasnoun{redelmeier1991assessing} test has to be below $p_{MSE,I}$.\\

Finally, to control for overfitting, the Aikaike information criterion (AIC) and variance inflation factors \cite{fox1992generalized} could be used to control for the number of variables included and multi-collinearity, respectively. An variable is added to a previously selected model only if the resulting AIC value is reduced and if variance inflation factors are below a threshold $v_{crit}$ that has to be defined by the user.\\

The tests described above are used to define the notion of an improved model. The criteria to be fulfilled are listed in Table \ref{tab:improved} below.
\begin{table}[!ht]
	\centering
	\begin{tabular}{ll} \hline
			Description & Quantification \\ \hline
			$\tilde{\beta}_c$ statistically significant & $p$-value of likelihood ratio test $< p_{lr,I}$ \\
			$\tilde{\beta}_c$ within theoretical expectation & Sign of $\tilde{\beta}_c$ matches expectation of statistician \\
			No multi-collinearity in $\mathbb{M}_c$ & Variance inflation factors $< v_{crit}$ \\
			$\mathbb{M}_c$ is well calibrated & $p$-value of \citeasnoun{spiegelhalter1986probabilistic} test $> p_{calib}$ \\
			$\mathbb{M}_c$ does not show overfitting & Akaike information criterion: AIC($\mathbb{M}_c$) $<$ AIC($\mathbb{M}$) \\
			$\mathbb{M}_c$ discriminates better than $\mathbb{M}$ & $p$-value of \citeasnoun{delong1988comparing} test $< p_{AUC,I}$ \\
			$\mathbb{M}_c$ is better calibrated than $\mathbb{M}$ & $p$-value of  \citeasnoun{redelmeier1991assessing} test $< p_{MSE,I}$ \\ \hline
	\end{tabular}
	\caption{\label{tab:improved}List of criteria the model $\mathbb{M}_c$ has to fulfill to be considered as improved over $\mathbb{M}$}
\end{table}

To make this part of the CSSLR algorithm applicable, a statistician has to define a table with expected signs of model coefficients. To give an example, when building a model for the creditworthiness of corporations, an analyst would expect that high profitability improves the creditworthiness (negative sign) and high debt reduces creditworthiness (positive sign). The expected impact of a variable on $P(I=1)$ should be clarified before the start of model building and verified whenever a new variable is added to a model. In situations where no expectation on the sign of a variable's coefficient could be formed, the sign check will be omitted. Besides that, values for the parameters $p_{lr,I}$, $p_{calib}$, $p_{AUC,I}$, $p_{MSE,I}$, and $v_{crit}$ have to be defined. Some fine-tuning of these parameters during a model selection process might be required to ensure that the algorithm does not select a too large number of models. This depends mostly on the data set and the number of bad observations.\\

Some of the criteria in Table \ref{tab:improved} are debatable. While most statisticians should agree on the first five criteria, some might prefer a weaker notion of improved. One could consider a model as improved if it either shows a significantly higher $AUC$ or a significantly lower $MSE$ and is not significantly weaker in the other measure. This would allow the selection of a wider range of models. This consideration illustrates that the CSSLR algorithm offers some flexibility and might be more difficult to parameterize compared to the simple forward and backward selection algorithms. However, its big advantage is that selected models will fulfill a much broader range of quality criteria.\\

Once a model $\mathbb{M}_c$ is identified as an improved model, it should be validated to ensure that the variables that have been included before $V_c$ still show the desired behavior. If one ore more variables no longer show a positive contribution to model performance, one might consider removing them from the model, i.e., to trim model $\mathbb{M}_c$. For this purpose an incremental analysis will be performed as outlined in Algorithm \ref{alg:trim}.

\begin{breakablealgorithm}\label{alg:trim}
	\caption{Incremental Analysis to Trim a Model}
	\begin{algorithmic}
		\For{$i=1,\ldots,m$}\Comment{Loop over all variables previously included in $\mathbb{M}_c$}
			\State Check sign $sgn$ of $\tilde{\beta_i}$ and value of $p_{lr}$
			\State Remove $V_i$ from $\mathbb{M}_c$, run the difference tests to compute $p_{AUC}$ and $p_{MSE}$
			\If{$sgn$ is wrong}
				\State Remove $V_i$ from $\mathbb{M}_c$
			\ElsIf{$p_{lr} > p_{lr,T}$ AND $p_{AUC} > p_{AUC,T}$ AND $p_{MSE} > p_{MSE,T}$}
				\State Remove $V_i$ from $\mathbb{M}_c$
			\EndIf
			\If{$V_i$ is removed}
				\State Stop the for-loop
				\State Rerun for-loop on the reduced model until no more trimming is needed
			\EndIf
		\EndFor
	\end{algorithmic}
\end{breakablealgorithm}

The idea of Algorithm \ref{alg:trim} is to provide an additional validation of model $\mathbb{M}_c$ before it is accepted as an improved model. Most importantly, it has to be ensured that the signs of model coefficients are still within expectations after including variable $V_c$. Furthermore, each variable $V_i$ should still have some positive contribution to the model, either by having a significant model coefficient, improving $AUC$, or improving $MSE$. Only if no positive contribution of $V_i$ to model $\mathbb{M}_c$ is visible, it should be removed. This process is controlled by three additional parameters $p_{lr,T}$, $p_{AUC,T}$, and $p_{MSE,T}$ that have to be defined when running the CSSLR algorithm.\\

After the search for improved models and their trimming, the outcome is not necessarily unique but there might be a set of candidate models. In the second part of the CSSLR algorithm, the candidate models in this set are compared in order to identifying a smaller subset of models that is superior in statistical terms. Only the smaller subset of superior models is kept and used as input in the next selection step. A starting point in this comparison is identifying the leading models. This is done by analyzing the two key dimensions of logistic regression models, discrimination and calibration. If there is a single model dominating in both dimensions, the leading model can be unique. If this is not the case, the number of leading models is two. The identification of leading models is described in Algorithm \ref{alg:leading}.

\begin{breakablealgorithm}\label{alg:leading}
	\caption{Determination of Leading Models}
	\begin{algorithmic}
		\State Determine model $\mathbb{M}_1$ with the largest $AUC$ value
		\State Determine model $\mathbb{M}_2$ with the smallest $MSE$ value
		\If{$\mathbb{M}_1 = \mathbb{M}_2$}
			\State Leading model is unique
		\Else
			\State Run the AUC and MSE difference tests and compute $p_{AUC}$ and $p_{MSE}$
			\If{$p_{AUC} < p_{AUC,E}$ AND $p_{MSE} > p_{MSE,E}$}
				\State Leading model is $\mathbb{M}_1$
			\ElsIf{$p_{AUC} > p_{AUC,E}$ AND $p_{MSE} < p_{MSE,E}$}
				\State Leading model is $\mathbb{M}_2$
			\Else
				\State Leading models are $\mathbb{M}_1$ and $\mathbb{M}_2$
			\EndIf		
		\EndIf
	\end{algorithmic}
\end{breakablealgorithm}

When comparing models $\mathbb{M}_1$ and $\mathbb{M}_2$ in Algorithm \ref{alg:leading}, the tests on difference in $AUC$ and $MSE$ are run. When there is a statistically significant difference in $AUC$ but not in $MSE$, model $\mathbb{M}_1$ is considered as superior and defined as the leading model. If it is the other way round, model $\mathbb{M}_2$ is dominating model $\mathbb{M}_1$. If both tests or none of the tests results in statistical significant outcomes, both models are considered as statistically equivalent and both models are kept in the list of candidate models for the next selection step. To decide on the statistical equivalence of two models, critical $p$-values $p_{AUC,E}$ and $p_{MSE,E}$ have to be defined before running the CSSLR algorithm.\\

When determining the leading models in Algorithm \ref{alg:leading}, the notion of equivalent models was introduced. These are models that cannot be rank-ordered in terms of discrimination and calibration, either because they are indistinguishable in both dimension, or because one model has the higher discriminative power and the second model the lower calibration error. This explains the final part of a selection step in Algorithm \ref{alg:select} where a model is compared with the leading models and kept in case it is equivalent or discarded, otherwise.\\

The stepwise selection is starting from a constant model. It adds variables one-by-one until either including additional variables does not lead to further improvements of selected models or a pre-defined maximum of selection steps is reached. In the remainder of this article, the performance of the CSSLR algorithm will be illustrated.

\section{Performance of CSSLR}\label{sec:performance}

The CSSLR algorithm is evaluated on multiple data sets generated by simulation. The starting point is a vector containing the good/bad indicator variables $I$. It contains $K$ good events coded by "0" and $K$ bad events represented by "1". On this data set, strong, weak and non-discriminating variables are created. The conditional distributions of strong variables $S_i$, weak variables $W_i$ and non-discriminating variables $R_i$ are given as
\begin{align*}
	S_i(I=0) & \sim N\left(\mu = \mu_1, \sigma = 1\right) \displaybreak[1] \\
	S_i(I=1) & \sim N\left(\mu = -\mu_1, \sigma = 1\right) \displaybreak[1] \\
	W_i(I=0) & \sim N\left(\mu = \mu_2, \sigma = 1\right) \displaybreak[1] \\
	W_i(I=1) & \sim N\left(\mu = -\mu_2, \sigma = 1\right) \displaybreak[1] \\
	R_i(I=0) & \sim N\left(\mu = 0, \sigma = 1\right) \displaybreak[1] \\
	R_i(I=1) & \sim N\left(\mu = 0, \sigma = 1\right)
\end{align*}
where $N$ is a normally distributed variable with expectation $\mu$ and standard deviation $\sigma$. Different values of $\mu_1 > \mu_2$ will be chosen to evaluate the performance of the selection algorithm. The generation of data is done with $K = 500$ and the number of random data sets generated for model selection is 1000 in each simulation run. \\

To illustrate the sensitivity of the CSSLR algorithm, four different sets of parameters are chosen to control the selection algorithm. They are displayed in Table \ref{tab:csslr:params} below. The final row of this table deserves more explanation. Model improvement is evaluated in the CSSLR algorithm mainly by using the AUC-test and the MSE-test. Two version are analyzed: First, a model is considered as improved over a reference model if one of these two tests indicates improvement and the second one indicates equivalence. In this case, it is sufficient to see improvement in one quantity while no deterioration is visible in the other. Second, in a more conservative version, a model is considered as improved if both AUC and MSE are improved significantly. In this case, the algorithm is expected to terminate earlier as it applies stricter criteria.\\ 

\begin{table}[ht]
\centering
\begin{tabular}{rcccccc}
  \hline
Parameter & CSSLR1a & CSSLR1b & CSSLR2a & CSSLR2b \\ \hline
 $p_{lr,I}$ & 5.0 & 5.0 & 5.0 & 5.0 \\
 $p_{calib}$ & 50.0 & 50.0 & 10.0 & 10.0 \\
 $v_{crit}$ & 5.0 & 5.0 & 5.0 & 5.0 \\
 $p_{AUC,I}$ & 5.0 & 5.0 & 10.0 & 10.0 \\
 $p_{MSE,I}$ & 5.0 & 5.0 & 10.0 & 10.0 \\
 $p_{AUC,T}$ & 2.5 & 2.5 & 2.5 & 2.5 \\
 $p_{MSE,T}$ & 2.5 & 2.5 & 2.5 & 2.5 \\
 $p_{AUC,E}$ & 5.0 & 5.0 & 10.0 & 10.0 \\
 $p_{MSE,E}$ & 5.0 & 5.0 & 10.0 & 10.0 \\
 Decision $I$ & AUC or MSE & AUC and MSE & AUC or MSE & AUC and MSE \\ \hline
\end{tabular}
\caption{Different sets of CSSLR parameters used to control the selection algorithm: $p_{lr,I}$ is the $p$-value of the model coefficient significance test in \%, $p_{calib}$ the $p$-value for the Spiegelhalter calibration test in \%, $v_{crit}$ the maximum acceptable variance inflation factor, $p_{AUC,I}$ the $p$-value of the AUC-test used to decide about model improvement in \%, and $p_{MSE,I}$ the $p$-value of the MSE-test used to decide about model improvement in \%. The $p$-values $p_{AUC,T}$ and $p_{MSE,T}$ are used to decide about model trimming and $p_{AUC,E}$ and $p_{MSE,E}$ to determine model equivalence. The row "Decision $I$" specifies the criteria used for the model improvement decision.}
\label{tab:csslr:params}
\end{table}

To see how CSSLR compares with existing methods, four alternative selection algorithms are applied. The first alternative is stepwise selection based on AIC which is implemented in the function stepAIC of the R package MASS \cite{ripley2013package}. As a second alternative, code was extracted from the R package My.stepwise \cite{ihscc2017package} to create a routine that selects variables based on $p$-values of model coefficient significance tests. The critical $p$-value was set to 5.0\% to be consistent with the parameterization of CSSLR in Table \ref{tab:csslr:params}. The final alternative are two versions of the LASSO taken from the R package glmnet \cite{hastie2021introduction}. The LASSO depends on a penalty parameter $\lambda$ in the estimation of the model equation. In glmnet, cross validation is used to suggest sensible choices for $\lambda$ based on the distribution of estimation errors. Lasso1 uses the optimal value of $\lambda$, $\lambda_o$, which minimizes the cross validation error while Lasso2 uses $\lambda_1 > \lambda_o$ which leads to a cross validation error of one standard deviation higher than the minimum value. This results in a more regularized version of the LASSO. Lasso2 will, therefore, in general lead to more parsimonious models than Lasso1. \\

\begin{table}[!ht]
\centering
\begin{tabular}{rrrrrrr} \hline
Method & $P_{s}$ & $A_{s}$ & $P_{w}$ & $A_{w}$ & $P_{nd}$ & $A_{nd}$ \\ \hline
 CSSLR1a & 100.00 & 3.00 &  99.70 & 2.39 & 1.50 & 1.00 \\ 
 CSSLR1b & 100.00 & 3.00 & 92.70 & 1.71 & 0.00 & NaN \\ 
 CSSLR2a & 100.00 & 3.00 & 100.00 & 2.72 & 4.90 & 1.02 \\ 
 CSSLR2b & 100.00 & 3.00 & 98.10 & 2.16 & 0.30 & 1.00 \\ 
 AIC & 100.00 & 3.00 & 100.00 & 3.00 & 93.60 & 2.85 \\ 
 Coeff & 100.00 & 3.00 & 100.00 & 2.98 & 57.10 & 1.51 \\ 
 Lasso1 & 100.00 & 3.00 & 100.00 & 3.00 &  99.90 & 7.31 \\
 Lasso2 & 100.00 & 3.00 & 100.00 & 3.00 & 58.60 & 2.04 \\ \hline
\end{tabular}
\caption{Results of automated model selection from a data set of 3 strong ($\mu=\pm 1$), 3 weak ($\mu=\pm 0.5$) and 14 nuisance variables. The methods evaluated are four versions of CSSLR, an AIC-based forward selection method, a forward selection based on coefficient $p$-values, and two version of the LASSO. $P_{s}$ / $P_{w}$ /$P_{nd}$ is the percentage of simulation runs where at least one strong / weak / non-discriminating variable was selected and $A_{s}$ / $A_{w}$ / $A_{nd}$ is the average number of strong / weak / non-discriminating variables selected conditional on the number of selected strong / weak / non-discriminating variables being at least one.}
\label{tab:data}
\end{table}

The first test uses three strong variables with $\mu_1 = \pm 1.0$, three weak variables with $\mu_2 = \pm 0.5$ and 14 nuisance variables resulting in a data set of 20 variables besides the response variable. To get an impression on the strength of these variables, note that $\mu =  \pm 1.0$ leads to variables with an $AUC$ of about 90\% while $\mu =  \pm 0.5$ creates variables with an $AUC$ of about 75\%. The results of the eight selection algorithms are displayed in Table \ref{tab:data}. For each method, the percentage of simulations is reported where at least one strong / weak / nuisance variable is selected. In addition, the average number of strong / weak / nuisance variables selected is computed conditional on being greater than zero. All methods select all strong variables in all scenarios. The differences are in selecting weak and nuisance variables. Overall, CSSLR is selecting more parsimonious models compared to the alternatives. There are multiple scenarios where CSSLR does not select weak variables, especially when "AUC and MSE" is used for deciding about model improvement. When "AUC or MSE" is used, only in 0.3\% of all scenarios CSSLR1a does not find weak variables while CSSLR2a always includes weak variables. Both CSSLR1a and CSSLR2a have a higher tendency of selecting non-discriminating variables where CSSLR2a performs worst with including non-discriminating variables in 4.9\% of all simulation runs. \\

Compared to CSSLR, the four reference methods select more variables on average. The two best performing methods Coeff and Lasso2 include nuisance variables in more than 50\% of all simulation runs which is substantially worse than all versions of CSSLR. Furthermore, the number of nuisance variables included is larger. While CSSLR when it selects nuisance variables mostly includes one variable only, the alternative selection methods in many cases select two or more. This makes a variable selection based on CSSLR more reliable since it does a better job in rejecting nuisance variables and includes variables only that have power in explaining the response variable.\\

\begin{table}[!ht]
	\centering
	\begin{tabular}{rrrrrrr} \hline
		Method & $P_{s}$ & $A_{s}$ & $P_{w}$ & $A_{w}$ & $P_{nd}$ & $A_{nd}$ \\ \hline
		CSSLR1a & 100.00 & 3.00 & 92.60 & 1.85 & 0.70 & 1.00 \\ 
		CSSLR1b & 100.00 & 3.00 & 87.00 & 1.72 & 0.30 & 1.00 \\ 
		CSSLR2a & 100.00 & 3.00 & 99.20 & 2.34 & 3.50 & 1.09 \\ 
		CSSLR2b & 100.00 & 3.00 & 97.60 & 2.18 & 2.20 & 1.05 \\ 
		AIC & 100.00 & 3.00 & 100.00 & 2.99 & 91.90 & 2.46 \\ 
		Coeff & 100.00 & 3.00 & 100.00 & 2.95 & 52.80 & 1.36 \\ 
		Lasso1 & 100.00 & 3.00 & 100.00 & 3.00 &  99.90 & 6.40 \\ 
		Lasso2 & 100.00 & 3.00 & 100.00 & 2.92 & 32.90 & 1.44 \\ \hline
	\end{tabular}
	\caption{Results of automated model selection from a data set of three strong ($\mu=\pm 0.3$), three weak ($\mu=\pm 0.15$) and 14 nuisance variables. The methods evaluated are four versions of CSSLR, an AIC-based forward selection method, a forward selection based on coefficient $p$-values, and two version of the LASSO. $P_{s}$ / $P_{w}$ /$P_{nd}$ is the percentage of simulation runs where at least one strong / weak / non-discriminating variable was selected and $A_{s}$ / $A_{w}$ / $A_{nd}$ is the average number of strong / weak / non-discriminating variables selected conditional on the number of selected strong / weak / non-discriminating variables being at least one.}
	\label{tab:data2}
\end{table}

In a second experiment, the strength of both strong and weak variables are reduced. The motivation is bringing these variables in terms of AUC closer to the nuisance variables and see whether CSSLR is still able to separate them. Here, $\mu_1 = \pm 0.30$ and $\mu_2 = \pm 0.15$ are used. These numbers roughly correspond to $AUC = 65\%$ and $AUC = 58\%$, respectively. The results are shown in Table \ref{tab:data2}. The results are comparable to Table \ref{tab:data}. Still all strong variables are selected while the percentage of CSSLR runs selecting weak variables is slightly decreased. However, the ability to identify nuisance variables is still strong and the percentage of scenarios where CSSLR selects nuisance variables is well below 5\% for all four parameterizations.\\

In the third test, the strong variables are removed from the data sets of the second experiment and replaced by nuisance variables resulting in data set where three variables have weak and 17 variables have no discriminatory power. The results are presented in Table \ref{tab:data3}. In this case, the percentages for selecting weak variables are increased for CSSLR compared to Table \ref{tab:data2} and the ability to reject nuisance variables remains strong. The four alternatives still select non-discriminating variables in too many scenarios. The best performing method is Lasso2 which selects nuisance variables in 32.9\% of all simulation runs which is still substantially higher than the numbers for CSSLR. \\

\begin{table}[!ht]
\centering
\begin{tabular}{rrrrr} \hline
Method & $P_{w}$ & $A_{w}$ & $P_{nd}$ & $A_{nd}$ \\ \hline
 CSSLR1a & 100.00 & 2.42 & 0.90 & 1.00 \\ 
 CSSLR1b & 98.80 & 2.13 & 0.00 & NaN \\ 
 CSSLR2a & 100.00 & 2.76 & 5.10 & 1.02 \\ 
 CSSLR2b & 100.00 & 2.54 & 1.10 & 1.00 \\ 
 AIC & 100.00 & 3.00 & 95.00 & 2.89 \\ 
 Coeff & 100.00 & 2.99 & 61.00 & 1.46 \\ 
 Lasso1 & 100.00 & 3.00 & 95.50 & 5.09 \\ 
 Lasso2 & 100.00 & 2.97 & 32.30 & 1.64 \\ 
 \hline
\end{tabular}
\caption{Results of automated model selection from a data set of 3 $\mu=\pm 0.15$ and 17 nuisance variables. The methods evaluated are four versions of CSSLR, an AIC-based forward selection method, a forward selection based on coefficient $p$-values, and two version of the LASSO. $P_{w}$ /$P_{nd}$ is the percentage of simulation runs where at least one weak / non-discriminating variable was selected and $A_{w}$ / $A_{nd}$ is the average number of weak / non-discriminating variables selected conditional on the number of selected strong / weak / non-discriminating variables being at least one.}
\label{tab:data3}
\end{table}

Finally, the test is run on data sets containing 20 non-discriminating variables. The correct behavior of a selection algorithm would be rejecting all variables and proposing a model with the constant only as independent variable. From Table \ref{tab:random} it can be seen that CSSLR is performing considerably better than  three of the reference methods while Lasso2 is comparable to CSSLR in terms of the number of scenarios where a nuisance variable is selected. When this happens, however, Lasso2 tends to select on average two nuisance variables while CSSLR selects one variable only. The best performing method is the most restrictive version of CSSLR, CSSLR2a, where in all scenarios the correct model with the constant is selected.\\

\begin{table}[!ht]
\centering
\begin{tabular}{rcc} \hline
Method & $P_{nd}$ & $A_{nd}$ \\ \hline
 CSSLR1a & 15.70 & 1.00 \\ 
 CSSLR1b & 0.00 & NaN \\ 
 CSSLR2a & 15.70 & 1.00 \\ 
 CSSLR2b & 2.00 & 1.00 \\ 
 AIC & 97.10 & 3.36 \\ 
 Coeff & 64.70 & 1.53 \\ 
 Lasso1 & 35.60 & 3.75 \\
 Lasso2 & 11.60 & 1.95 \\ \hline
\end{tabular}
\caption{Results of automated model selection from a data set of 20 nuisance variables. The methods evaluated are four versions of CSSLR, an AIC-based forward selection method, a forward selection based on coefficient $p$-values, and two version of the LASSO. $P_{nd}$ is the percentage of simulation runs where at least one non-discriminating variable was selected and $A_{nd}$ is the average number of non-discriminating variables selected conditional on the number of selected non-discriminating variables is at least one.}
\label{tab:random}
\end{table}

\section{Conclusions}

In this article, a comprehensive stepwise model selection algorithm for logistic regression, CSSLR, was proposed. In contrast to existing model selection methods, CSSLR is less generic and tailored to logistic regression by focusing on its two key dimensions, discriminatory power and calibration. A model's discriminatory power is measured by AUC while calibration is measured by MSE. Starting from a model with the constant only, new variables are added one-by-one if they fulfill basic requirements like significance tests for model coefficients and low variance inflation factors and, in addition, pass tests on improvement of AUC and MSE.  The outcome of the selection process may not be a single model but multiple models that could be considered as equivalent in terms of AUC and MSE.\\

In a simulation study CSSLR was compared with model selection based on AIC, $p$-values of significance tests for model coefficients and two versions of the LASSO. It was demonstrated that CSSLR is superior to these methods in terms of selecting meaningful variables while at the same time rejecting nuisance variables. In all experiments the percentages of simulations where CSSLR selected nuisance variables was substantially lower than for the tested alternatives. This gives some confidence that in practical applications, CSSLR will lead to parsimonious models selecting the most important variables only while variables representing data noise will most likely be filtered out.\\

The superior performance of CSSLR comes at a price. Compared to the alternatives analyzed in this article, the parameterization of CSSLR is more complex since thresholds for multiple $p$-values have to be defined. Furthermore, for the variable selection on a rich data set it might be necessary to perform thousands of regression model estimations and statistical tests which results in substantially higher computational times. While the routines in the R packages MASS and glmnet are computationally efficient and deliver solutions within seconds, CSSLR might take minutes or for large datasets even hours until the selection process is completed. Despite of this, CSSLR should still save a data analyst a lot of time in analyzing a model estimation problem because it gives transparent results of every step in the selection process. It shows why certain models have been rejected or selected and documents the full process of arriving at the final models.\\

Finally, it should be noted that the high-level variable selection method outlined in Algorithm \ref{alg:select} is generic and did not use any properties of logistic regression. This means that it should be possible to improve model selection algorithms utilizing Algorithm \ref{alg:select} for other classes of statistical models by tailoring the notion of improved and equivalent to their characteristics. Exploring variable selection for other model families is beyond the scope of this article and left for future research.

\section{Appendix}

The code of CSSLR is written in R. To replicate the results of this article and perform own tests of the selection algorithm, the code can be accessed on \newline\url{https://github.com/berndengelmann/CSSLR}. The easiest way of installing the package is using the command \texttt{devtools::install\_github("berndengelmann/CSSLR")}. After installing the package, test scripts could be found in a subfolder \texttt{Tests} in the folder where the package is installed. The script \texttt{ModelSelectionSimulation\_Article.R} allows the replication of the tables presented in this article.

\bibliographystyle{agsm}
\bibliography{bib}

\end{document}